\UseRawInputEncoding
\documentclass[aps,prl,reprint,showpacs,superscriptaddress]{revtex4-1}

\usepackage{graphicx}
\usepackage{color}
\usepackage[colorlinks, linkcolor=blue, anchorcolor=blue, citecolor=blue, urlcolor=blue]{hyperref}
\usepackage{multirow}
\usepackage{amsmath}

\begin{document}


\title{Anomalous ferromagnetic behavior in the orthorhombic Li$_3$Co$_2$SbO$_6$}



\author{Qianhui Duan}
\thanks{These authors contribute equally.}
\affiliation{School of Physics and Hangzhou Key Laboratory of Quantum Matters,
Hangzhou Normal University, Hangzhou 311121, China}
\affiliation{Neutron Science Platform, Songshan Lake Materials Laboratory, Dongguan, Guangdong
523808, China}

\author{Huanpeng Bu}
\thanks{These authors contribute equally.}
\affiliation{Neutron Science Platform, Songshan Lake Materials Laboratory, Dongguan, Guangdong
523808, China}

\author{Vladimir Pomjakushin}
\affiliation{Laboratory for Neutron Scattering and Imaging LNS, Paul Scherrer Institute, CH-5232
Villigen PSI, Switzerland}

\author{Hubertus Luetkens}
\affiliation{Laboratory for Muon Spin Spectroscopy, Paul Scherrer Institute, CH-5232 Villigen PSI,
Switzerland}

\author{Yuke Li}
\email{yklee@hznu.edu.cn}
\affiliation{School of Physics and Hangzhou Key Laboratory of Quantum Matters,
Hangzhou Normal University, Hangzhou 311121, China}

\author{Jinkui Zhao}
\affiliation{Neutron Science Platform, Songshan Lake Materials Laboratory, Dongguan, Guangdong
523808, China}

\author{Jason S. Gardner}
\affiliation{Material Science \& Technology Division, Oak Ridge National Laboratory, Oak Ridge,
Tennessee 37831, USA}

\author{Hanjie Guo}
\email{hjguo@sslab.org.cn}
\affiliation{Neutron Science Platform, Songshan Lake Materials Laboratory, Dongguan, Guangdong
523808, China}



\date{\today}

\begin{abstract}
  Monoclinic Li$_3$Co$_2$SbO$_6$ has been proposed as a Kitaev spin liquid candidate and
  investigated intensively, whereas the properties of its polymorph, the orthorhombic phase, is
  less known. Here we report the magnetic properties of the orthorhombic Li$_3$Co$_2$SbO$_6$ as
  revealed by dc and ac magnetic susceptibility, muon spin relaxation ($\mu$SR) and neutron
  diffraction measurements. Successive magnetic transitions at (115, 89 and 71) K were observed in
  the low field dc susceptibility measurements.  The transitions below $T_N$ (= 115 K), are
  suppressed in higher applied fields. However, zero field, ac susceptibility measurements reveals
  distinct frequency independent  transitions at about (114, 107, 97, 79 and 71) K. A long range
  magnetic ordered state was confirmed by specific heat, $\mu$SR and neutron diffraction
  measurements, all indicating a single transition at about 115 K. The discrepancy between
  different measurements is attributed to possible stacking faults and/or local disorders of the
  ferromagnetic zig-zag chains, resulting in ferromagnetic boundaries within the overall
  antiferromagnetic matrix.
\end{abstract}


\maketitle


\section{Introduction}
Co$^{2+}$ based honeycomb lattice materials are of great interests due to the possible realization
of the Kitaev model
\cite{Liu2018,Sano2018,Xiao2019,Yan2019,Liu2020,Songvilay2020,Yao2020,Lin2021,Lef2016,Bera2017}. The
3\textit{d} electrons are more localized compared to those of the 4\textit{d} (Ru) or 5\textit{d}
(Ir) systems, and could suppress the undesired long range interactions beyond the nearest neighbors.
Moreover, the extra e$_g$-electrons introduce more interaction channels which can be tuned to
effectively quash the non-Kitaev terms in the Hamiltonian \cite{Liu2020}. Such candidates include
\textit{A}$_3$Co$_2$SbO$_6$ and \textit{A}$_2$Co$_2$TeO$_6$ (\textit{A} = alkali metals).
Among these, the Li$_3$Co$_2$SbO$_6$ has two polymorphs, one crystalizes in the monoclinic structure
with the Co honeycomb lattice, and the other in the orthorhombic structure \cite{Brown2019}. In the
monoclinic phase, the Co-O-Co angle is close to 90$^{\circ}$, making this material a better
approximation to the Kitaev model than the well studied RuCl$_3$ \cite{Cao2016} or Na$_2$IrO$_3$
\cite{Choi2012}. Recent neutron powder diffraction measurements on the monoclinic
Li$_3$Co$_2$SbO$_6$ confirm the formation of ferromagnetic honeycomb layers which stack
antiferromagnetically along the $\it c$-direction \cite{Vivanco2020}.
On the other hand, the orthorhombic phase is less investigated.

The main feature of the crystal structure of the orthorhombic Li$_3$Co$_2$SbO$_6$ can be viewed as a
stacking of Co-O zig-zag chains along the \textit{c}-axis. The chains, from layer to layer, run
along the [110] and [1-10] directions alternatively, as shown in Fig. \ref{structure}(a).
The Co ions along the chain are connected via edge-shared CoO$_6$ octahedra, while the interchain Co
ions are connected via corner-shared CoO$_6$ octahedra.
The intra- and interchain Co-O-Co angles are about 92$^{\circ}$ and 170$^{\circ}$, presumably
leading to antiferromagnetic and ferromagnetic interactions, respectively.
Previous studies by Brown \textit{et al.} have shown that the sample has three transitions at 113,
80 and 60 K, as revealed by dc magnetic susceptibility measurements. Similar multiple transitions in
the dc magnetic susceptibility below $T_N$ have been observed in the honeycomb phase such as
Na$_2$Co$_2$TeO$_6$ \cite{Bera2017,Xiao2019}. Also, the honeycomb phase of Li$_3$Co$_2$SbO$_6$
clearly shows a bifurcation at $\sim$75 K, well above the antiferromagnetic transition temperature
$T_N$ = 14 K \cite{Brown2019}. However, the nature of these multiple transitions remains unclear.
Brown \textit{et al}. have also performed neutron powder diffraction measurements on the title
compound, but unfortunately, failed to unravel the magnetic structure of the ordered state
\cite{Brown2019}. These findings demonstrate the rich properties in the orthorhombic phase, which
deserves further investigations.

In order to elucidate the magnetic ground state and the origin of the successive transitions, we
have performed detailed dc and ac magnetic susceptibility, together with muon spin relaxation
($\mu$SR) and neutron diffraction measurements, which enable us to reveal the underlying magnetic
ground state from a microscopic aspect. The most surprising result is the observation of multiple
sharp ferromagnetic transitions from the ac susceptibility measurement, which have almost no
correspondence in the dc magnetic susceptibility, nor in the $\mu$SR or neutron measurements. We
propose that these anomalies most likely originate from the stacking faults and/or local disorders
of the zig-zag chains, thus, forming ferromagnetic regions with small volume fraction that is
difficult to be detected by other techniques, but can not conclusively rule out other possibilities,
like charge fluctuations, spin stripes, or cluster glass dynamics.

\section{Experimental Section}

Polycrystalline samples of orthorhombic Li$_3$Co$_2$SbO$_6$ was prepared by a conventional solid
state reaction method. Li$_2$CO$_3$ (99.99\%) was dried at 120$^{\circ}$C for 4 hours prior to the
reaction. It was then mixed with Sb$_2$O$_3$ (99.99\%) in the appropriate ratio to form the
Li$_3$SbO$_4$ precursor, which was then mixed with stoichiometric CoO powders and ground thoroughly
in an agate mortar, pressed into pellet and calcined between 1100 and 1150$^{\circ}$C for 24 hours
with several intermediate grindings. The phase purity was confirmed both by laboratory x-ray
diffraction (XRD) and neutron powder diffraction (NPD) measurements.

The dc and ac magnetic susceptibility measurements were performed using the vibrating sample
magnetometer (VSM) and ACMS-II options, respectively, of the Physical Property Measurement System
(PPMS DynaCool, Quantum Design). For the ac susceptibility measurement, excitation fields of 5 Oe in
amplitude are superimposed to various dc bias fields (including zero field). The heat capacity was
measured using the relaxation method in the PPMS.

Muon spin relaxation measurements were performed on the GPS spectrometer at the Paul Scherrer
Institute (PSI), Villigen, Switzerland. Nearly 100\% polarized muons were injected into the sample
and the decayed positrons, which are emitted preferentially along the muon spin direction, were
detected. The asymmetry is defined as A(t) = [N(t) - $\alpha$B(t)]/[N(t) + $\alpha$B(t)], where N(t)
and B(t) are the number of positrons arrived at the forward and backward detectors at time t, and
the parameter $\alpha$ reflects the relative counting efficiencies of the two detectors. Here, the
forward and backward detectors are referred to the ones located at the top and down positions,
respectively, with the muon spin polarization being rotated about 45$^{\circ}$ with respect to the
beam. The data was analyzed using the Musrfit software.\cite{musrfit}

Neutron powder diffraction measurements were carried out on the HRPT diffractometer at PSI.
Approximate 10 g of the samples with natural Li were loaded into a vanadium can. Neutron wavelengths
of 1.15~\AA~and 1.89~\AA~were used for the nuclear and magnetic structure refinement, respectively.
The neutron data was analyzed using the FullProf software suite.\cite{Fullprof}

\section{Results}

\subsection{nuclear structure}

\begin{figure}
\centering
\includegraphics[width=0.8\columnwidth]{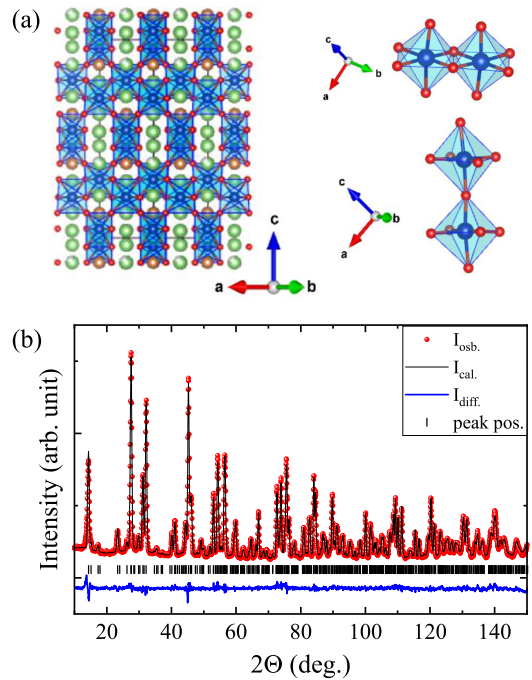}
\caption{(Color online) (a) Crystal structure of the orthorhombic Li$_3$Co$_2$SbO$_6$. Blue - Co,
red - O, green - Li, yellow - Sb. The local structure of edge-sharing and corner-sharing CoO$_6$
octahedra are also shown on the right side. (b) Rietveld refined pattern from Li$_3$Co$_2$SbO$_6$.
The data was collected at 130 K with neutron wavelength of 1.15 \AA.}
\label{structure}
\end{figure}

The synthesized sample crystallizes in the orthorhombic phase with space group $Fddd$. Since the
laboratory XRD is not sensitive to the light element Li, the nuclear structure refinement was
conducted against the NPD pattern collected at 130 K, as shown in Fig. \ref{structure}(b). Assuming
a full occupancy at each site, it was found that the isotropic thermal parameter, $B_{iso}$, of Li2
at the 8\textit{b} site is unusually large. This is consistent with a previous study, where it was
assumed that a small amount of Co ions replace the Li ions at this site \cite{Brown2019}. However,
in this case, some of the Co ions will possess the rare valence state of 1+. Therefore, a Li vacancy
at the Li2 site is assumed in our refinement, which is reasonable considering the volatile nature of
Li at high temperatures. This naturally leads to a mixing of the more common Co$^{2+}$ and Co$^{3+}$
valence states. The refined structural parameters are summarized in the Tab. \ref{cry}. It was
determined that there is a small Li deficiency of amount, $\sim$ 7\%, in our sample. In other words,
a maximum of about 7\% of the Co ions are in the 3+ valence state in order to keep charge
neutrality.

\begin{table}
\caption{Structural parameters of the orthorhombic Li$_3$Co$_2$SbO$_6$ obtained from Rietveld
refinement on the neutron data collected at 130 K with a neutron wavelength of 1.15~\AA. The space
group is \textit{Fddd} (No. 70). The lattice parameters are \textit{a} = 5.92526(4)~\AA, \textit{b}
= 8.68376(6)~\AA, \textit{c} = 17.91232(11)~\AA. The \textit{R}-values amount to $R_p$ = 8.39 \% and
$R_{wp}$ = 9.24\%. The fractional coordinations, isotropic displacement parameters $B_{iso}$, and
the occupancies are presented.\label{cry}}
\begin{tabular}{cccccc}
atom & x & y & z & $B_{iso}$ (\AA$^{2}$) & occ.  \\
\hline
 Li1 (16\textit{g}) &   0.12500     &  0.62500      &  0.28578(35) &  1.064(98)  & 1        \\
 Li2 (8\textit{b})  &   0.12500     &  0.62500      &  0.12500     &  0.887(165) & 0.79(2)  \\
 Co  (16\textit{g}) &   0.12500     &  0.12500      &  0.29475(21) &  0.263(46)  & 1        \\
 Sb  (8\textit{a})  &   0.12500     &  0.12500      &  0.12500     &  0.294(28)  & 1        \\
 O1  (16\textit{g}) &   0.12500     &  0.35378(13)  &  0.12500     &  0.417(20)  & 1        \\
 O2  (32\textit{h}) &   0.11178(14) &  0.37007(18)  &  0.29559(6)  &  0.314(10)  & 1         \\
\end{tabular}
\end{table}

\subsection{magnetic susceptibility}
\begin{figure}
\centering
\includegraphics[width=1\columnwidth]{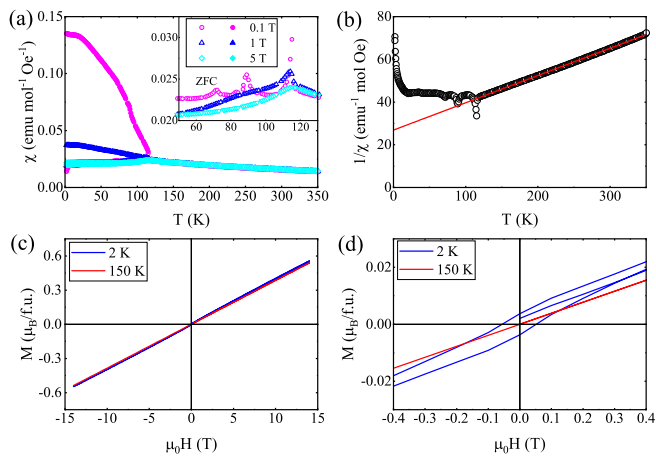}
\caption{(Color online) (a) Temperature dependence of the dc magnetic susceptibility measured in the
zero-field-cooled (ZFC) and field-cooled (FC) modes, represented by open and closed marks,
respectively. The inset highlights the ZFC curve around $T_N$. (b) Temperature dependence of the
inverse magnetic susceptibility $\chi^{-1}$ measured with $H$ = 1000 Oe, and a fit to the
Curie-Weiss law. (c) Isothermal magnetization measurements at various temperatures. (d) A closer
look up of the hysteresis loop around the zero field.}
\label{MT}
\end{figure}

Temperature dependence of the dc magnetic susceptibility is presented in Fig. \ref{MT}(a). A
bifurcation can be seen below $T_N \sim$ 115 K for the zero-field-cooled (ZFC) and field-cooled (FC)
curves, which is most apparent for the 0.1 T data, and gradually suppressed with increasing fields.
At 5 T, the two curves almost overlap. A careful inspection of the ZFC curve, as shown in the inset
of Fig. \ref{MT}(a), reveals more complex behavior below $T_N$. In addition to the peak at $\sim$115
K, two small peaks appear at $\sim$89 and $\sim$71 K, which are also suppressed by the magnetic
field, and already disappear at 1 T. A Curie-Weiss fit, $\chi^{-1} = (T-\theta_{CW})/C$, to the 0.1
T ZFC curve above 150 K resulting in an effective moment of 5.60 $\mu_B$/Co and Curie-Weiss
temperature $\theta_{CW}$ of -210 K. The large effective moment indicates that the Co$^{2+}$ ions
are in the high spin state (S = 3/2) with substantial orbital moment \cite{Guo2021}. The negative
$\theta_{CW}$ is suggestive of the dominant antiferromagnetic interactions among Co ions. The
bifurcation between the ZFC and FC curves suggests the existence of ferromagnetic components on top
of the antiferromagnetic ground state, which is further verified by the isothermal hysteresis
measurements. As shown in Fig. \ref{MT}(c), the M vs. H curves are almost linear down to the lowest
temperature, as expected for an antiferromagnet. However, a closer lookup of the curve in the low
field region clearly show hysteresis effect, consistent with the M vs. T measurements.

\begin{figure}
\centering
\includegraphics[width=0.9\columnwidth]{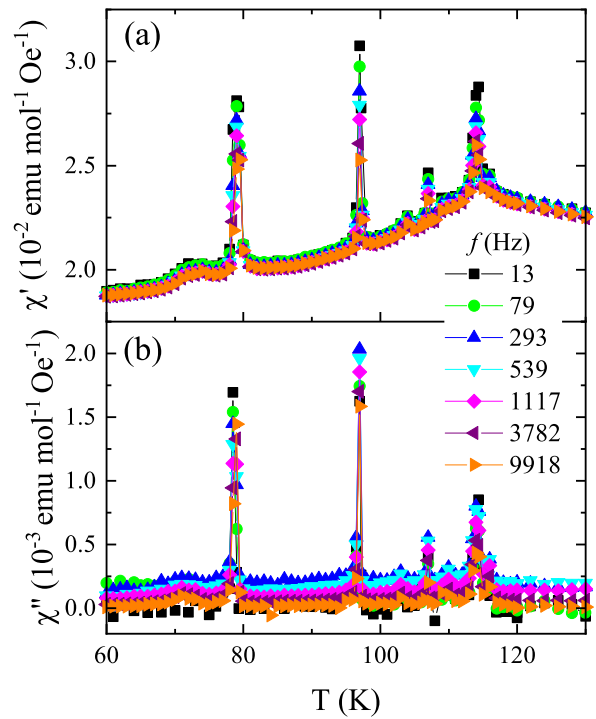}
\caption{(Color online) Temperature dependence of (a) the real $\chi'$ and (b) the imaginary
$\chi''$ components of the ac susceptibility measured at various frequencies and zero dc bias
field.}
\label{ac_zf}
\end{figure}

\begin{figure}
\centering
\includegraphics[width=0.9\columnwidth]{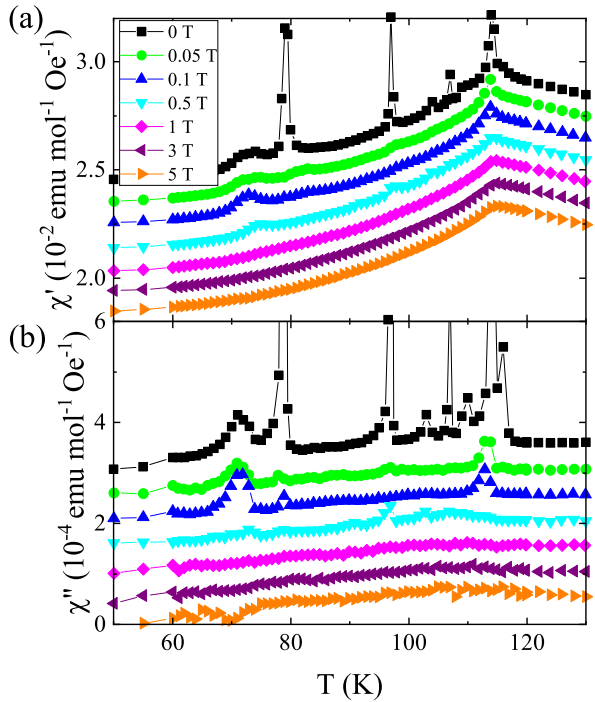}
\caption{(Color online) Temperature dependence of (a) the real $\chi'$ and (b) the imaginary
$\chi''$ components of the ac susceptibility measured with frequency = 3782 Hz and various dc bias
fields. The data has been shifted vertically for clarity.}
\label{ac_f}
\end{figure}

In order to further investigate the nature of the successive transitions below $T_N$, a series of ac
susceptibility measurements have been performed. The samples used for the dc and ac susceptibility
measurements shown here are from the same batch, but not exactly the same ones. Those using exactly
the same sample for the dc and ac susceptibility measurements are shown in the Supplementary
Materials (SM) Fig. S1. Figure \ref{ac_zf} show the temperature dependence of the real ($\chi'$) and
imaginary ($\chi''$) components of the ac susceptibility with various driven frequencies and no dc
bias field. Remarkably, at least four sharp peaks (FWHM less than 1 K) can be observed at $\sim$114,
107, 97 and 79 K both in $\chi'$ and $\chi''$. The sharp-peak positions are independent of the
frequencies, ruling out the formation of spin glass state \cite{Guo2016}. Combined with the presence
of nonzero $\chi''$, these results suggest the presence of the ferromagnetic components.
Surprisingly, the peaks below $T_N$ have no correlation with those observed below $T_N$ (=~115 K) in
the dc susceptibility measurements ($\sim$89~K and 71~K), see also Fig. S1 in SM. A more careful
inspection of the ac susceptibility data reveals a weak, broad peak at 73 K in $\chi'$ (71 K in
$\chi''$) in Fig. \ref{ac_f}, which can be associated with the low temperature peak (71 K) observed
in the dc measurement. However, the 89 K peak is still invisible in the ac susceptibility.

Figure \ref{ac_f} shows the dc field dependence of the ac susceptibility at a fixed frequency of
3782 Hz. The sharp peaks at (107, 97 and 79) K are profoundly affected by dc fields, and are almost
completely suppressed with a dc field of 0.05 T, whereas the 71 K peak persists to 0.1 T. Above 0.5
T, only the signature of $T_N$ survives, which behaves as a kink in $\chi'$, has no anomaly in
$\chi''$, and does not change in fields up to 5 T, consistent with an antiferromagnetic transition.

\subsection{specific heat}
\begin{figure}
\centering
\includegraphics[width=0.9\columnwidth]{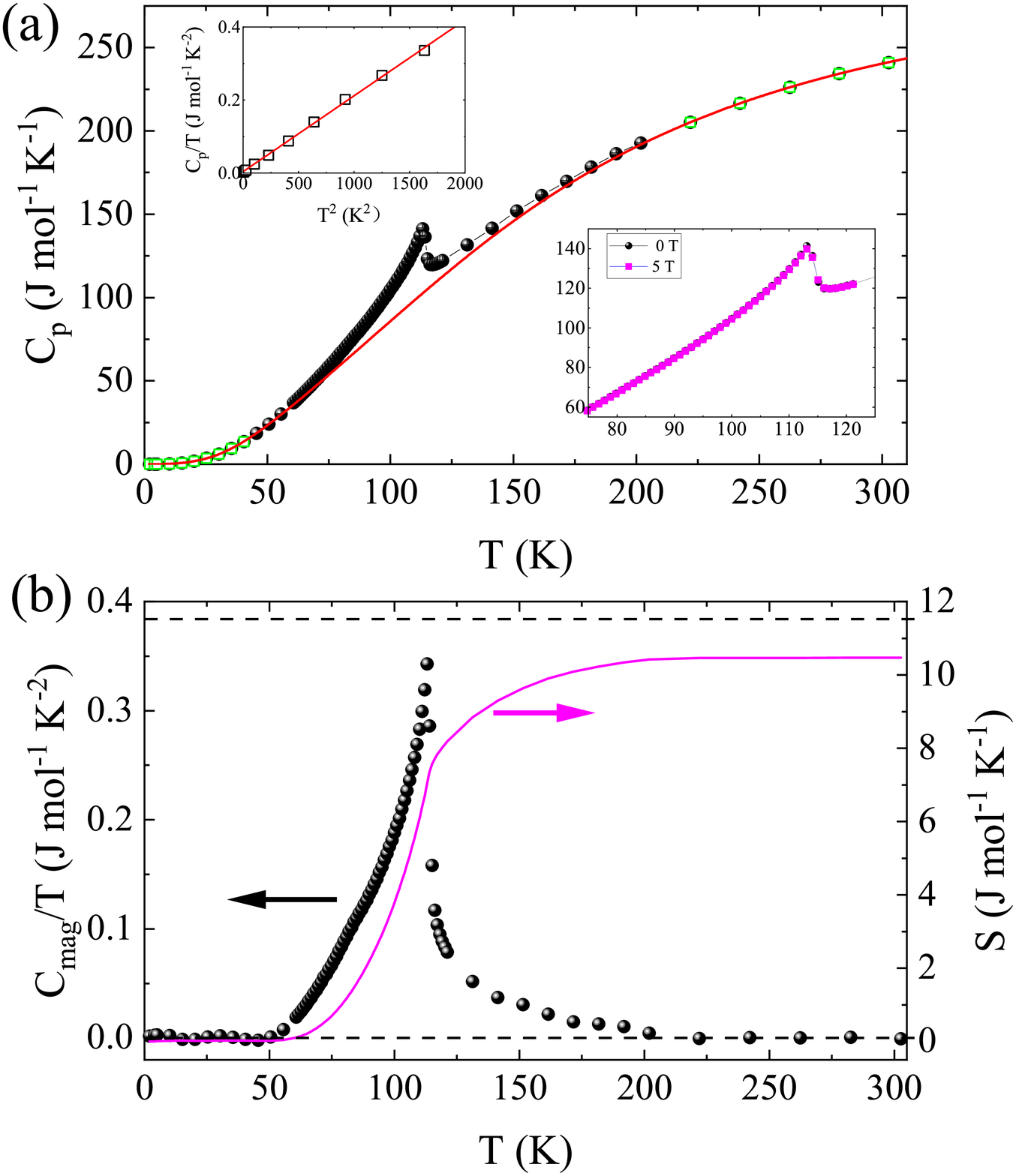}
\caption{(Color online) (a) Temperature dependence of the specific heat. The green data points are
used for the estimation of the lattice contributions, see the text for details. The red curve is the
fitting result according to the Debye + Einstein model. The left inset shows the low temperature
region and a fit with C$_p$ = $\gamma$T + $\beta$T$^3$. The right inset compares the heat capacity
measured both in zero and 5 T. (b) Magnetic heat capacity $C_{mag}$/T obtained by subtracting the
calculated phonon contribution from the total heat capacity, and the magnetic entropy obtained by
integration of $C_{mag}$/T. The upper dash line indicates the expected entropy value of 2$R$ln2 (the
factor 2 comes from 2 Co ions in the chemical formula).}
\label{HC}
\end{figure}

The specific heat, $C_p$, is presented in Fig. \ref{HC}(a). A $\lambda$-shaped transition can be
observed at $T_N$ = 115 K. No further anomaly is observed at lower temperatures. The transition at
115 K is very robust to magnetic field, no obvious changes can be observed at 5 T, as shown in the
right inset and consistent with the ac susceptibility. The low temperature ($<$ 40 K) region is
described well  by $C_p$(T)/T = $\gamma$ +$\beta$T$^2$, yielding a negligibly small $\gamma$ of 3(2)
mJ/mol-K$^2$, and $\beta$ of 2.08(2) $\times$ 10$^{-4}$ J/mol-K$^{4}$, indicating that the low
temperature specific heat is dominated by phonon contributions. The Debye temperature $\theta_D$
extracted from $(12\pi^{4}rR/5\beta)^{1/3}$, where \textit{r} is the number of atoms in the chemical
formula and $R$ is the ideal gas constant, amounts to 482 K. In order to extract the magnetic
contributions from the total heat capacity, the phonon contributions are estimated from a fit with a
combination of the Debye and Einstein model,
\begin{equation}\label{DE}
\begin{split}
  C_p(T) = & x_D\cdot9rR(T/\theta_D)^3\int_0^{\theta_D/T} x^4\mathrm{exp}(x)/[\mathrm{exp}(x)-1]^2dx
  + \\
         & (1-x_D)\cdot3rR(\theta_E/T)^2\mathrm{exp}(\theta_E/T)/[\mathrm{exp}(\theta_E/T)-1]^2,
\end{split}
\end{equation}
using the high temperature ($>$ 220 K) and low temperature ($<$ 40 K) data points marked as green
circles in Fig. \ref{HC}. The best fit results in about $x_D$ = 42\% weight to the Debye term. The
obtained Debye temperature, $\theta_D$, amounts to 356 K, in reasonable agreement with the one
extracted from the low temperature fitting alone, while the Einstein temperature $\theta_E$ is 620
K.
The magnetic contribution, $C_{mag}$, is thus obtained by subtracting the fitted phonon contribution
from the total heat capacity, and the magnetic entropy is then calculated by integrating
$C_{mag}/T$, as shown in Fig. \ref{HC}. The calculated release of entropy is about 10.5 J/mol-K,
which is close to the value expected for an effective spin s = 1/2 state [2\textit{R}ln(2s+1) =
11.52 J/mol-K].

\subsection{muon spin relaxation}
\begin{figure}
\centering
\includegraphics[width=0.8\columnwidth]{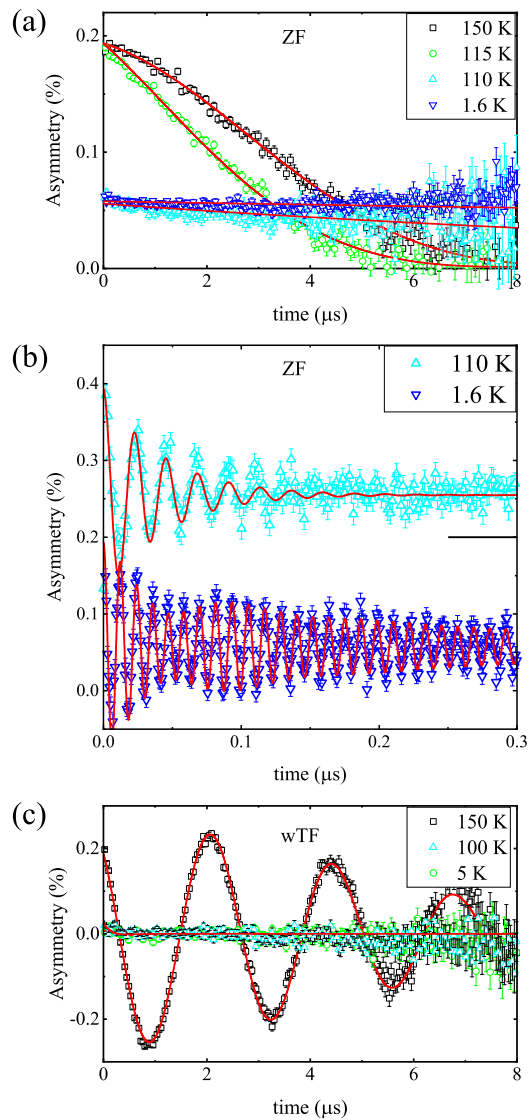}
\caption{(Color online) (a) Zero-field $\mu$SR spectra measured at various temperatures. (b) A
closer look up of the early time region showing the muon spin precessions at low temperatures. (c)
Weak transverse field (wTF) spectra at various temperatures. Solid lines represent the fits as
described in the text.}
\label{muSR_spectrum}
\end{figure}

In order to unravel the origin of the enigmatic magnetic behavior observed in the bulk, static and
dynamic susceptibility measurements, we have performed local-probed $\mu$SR measurements. Figure
\ref{muSR_spectrum}(a) shows the ZF-$\mu$SR time spectra measured at various temperatures. At 150 K,
the spectrum exhibits a gaussian-shaped depolarization behavior, indicating an electronic
paramagnetic state, and the depolarization is mainly caused by the nuclear moments. With a decrease
in temperature, the initial asymmetry drops quickly, showing a dramatic change of the spectra around
$T_N$ = 115 K. In the early time regions, Fig. \ref{muSR_spectrum}(b) exhibits clear muon spin
precessions at low temperatures. At the base temperature (1.6 K), two independent components are
required to describe the oscillating spectra creating a beating, however, by  $T_N$, the two
frequencies become comparable and merge into one frequency (see the 110 K spectrum). In order to
extract useful parameters from the spectra, a relaxation function
\begin{equation}\label{eq1}
   A(t) = A_0 G_{KT}(t)\mathrm{exp}(-\lambda_3 t)
\end{equation}
was fit to the spectra above 110 K, where $A_0$ is the initial asymmetry at time zero, $G_{KT}(t)$
the gaussian Kubo-Toyabe function \cite{Guo2013}, and $\lambda_3$ the relaxation rate caused by
electronic spins. At low temperatures, a three-component function
\begin{equation}
  \begin{split}
   A(t) = & A_1 \mathrm{cos}(\gamma_\mu B_{1}t)\mathrm{exp}(-\lambda_1 t) + A_2
   \mathrm{cos}(\gamma_\mu B_{2}t)\mathrm{exp}(-\lambda_2 t) \\
          & + A_3 \mathrm{exp}(-\lambda_3 t)
  \end{split}
\end{equation}
was used to describe the spectra, where $A_i$ is the amplitude of each component with the constraint
$\sum_i A_i = A_0$, $B_i$ the internal field at the muon site, $\gamma_\mu/2\pi$ = 13.55 MHz/kOe the
gyromagnetic ratio of muon, and $\lambda_i$ is the corresponding relaxation rate for each component.

\begin{figure}
\centering
\includegraphics[width=1\columnwidth]{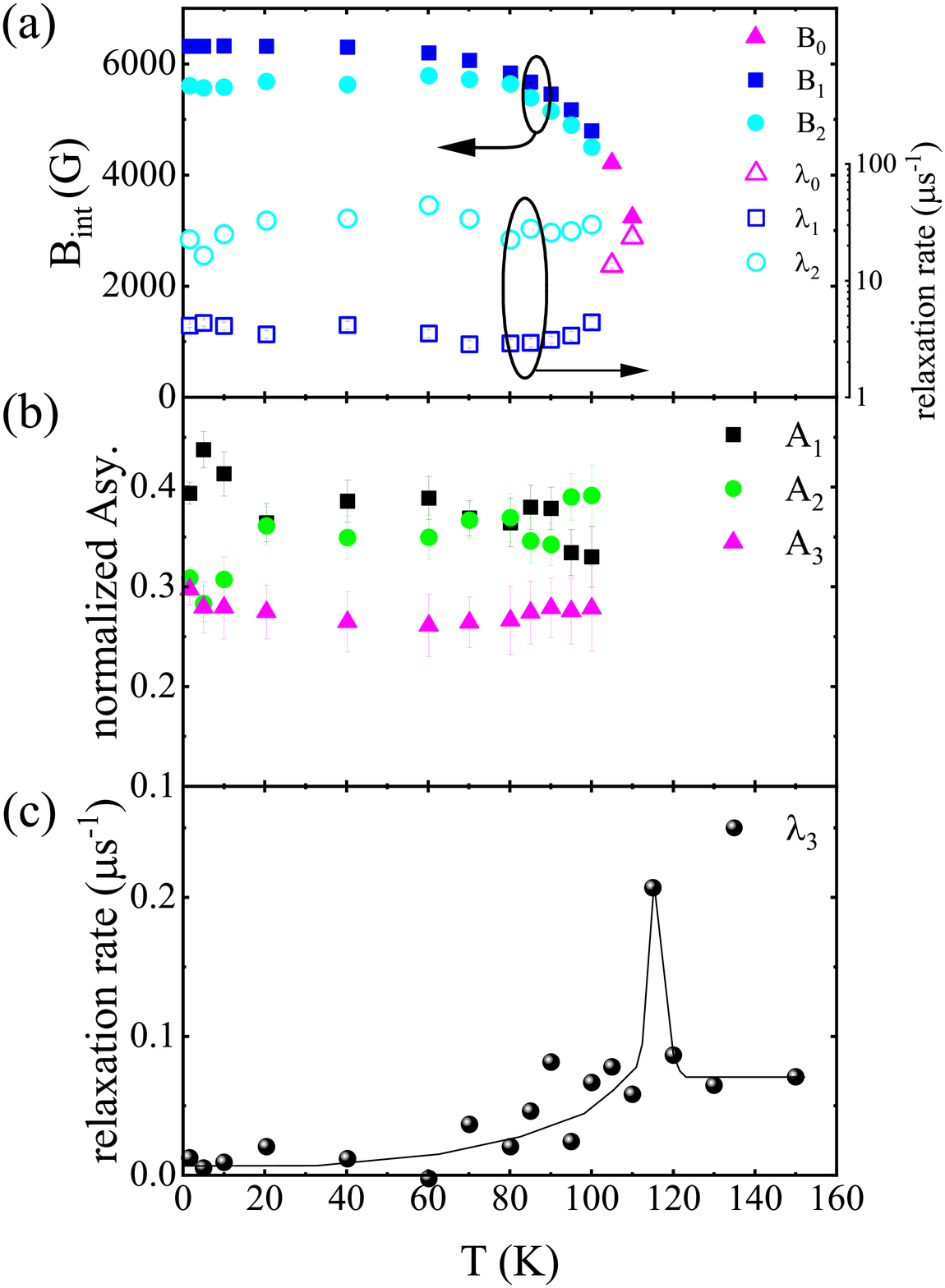}
\caption{(Color online) Temperature dependence of (a) the internal fields and corresponding
relaxation rates, (b) the amplitude of different components, and (c) the slow relaxation rate. The
solid line is the guide to the eyes.}
\label{muSR_parameter}
\end{figure}

The temperature dependence of the extracted parameters are shown in Fig. \ref{muSR_parameter}. Two
internal fields can be resolved at the base temperature and these remain almost constant until
$\sim$60 K. With the increase in the temperature, the difference gets smaller, and becomes
indistinguishable around $T_N$, denoted as $B_0$ in Fig. \ref{muSR_parameter}(a). It is worth noting
that the depolarization rate $\lambda_2$ is almost one order larger than $\lambda_1$, indicating a
much broader distribution for the $B_2$ component. The amplitude for these two components, $A_1$ and
$A_2$, are comparable over the whole temperature range. On the other hand, the fitted $A_3$
component, which reflects the portion of muons with the spin parallel to the internal fields, is
$\sim$ 0.3 at temperatures below 100 K, indicating that nearly 100\% of the spin system is static
\cite{note}. This is further corroborated from the weak transverse field (wTF) measurement as shown
in Fig. \ref{muSR_spectrum}(b). The paramagnetic volume fraction is proportional to the oscillation
amplitude, and it is clearly seen that the oscillation amplitude is almost absent below 100 K.
Finally, the magnetic transition temperature can also be inferred from the divergent behavior of
$\lambda_3$, which reflects the critical slowing down of the electronic spins, and a sharp peak can
be observed at 115 K, as shown in Fig. \ref{muSR_parameter}(c), consistent with the heat capacity
and high temperature magnetic transition.

\subsection{neutron powder diffraction}

\begin{figure}
\centering
\includegraphics[width=0.8\columnwidth]{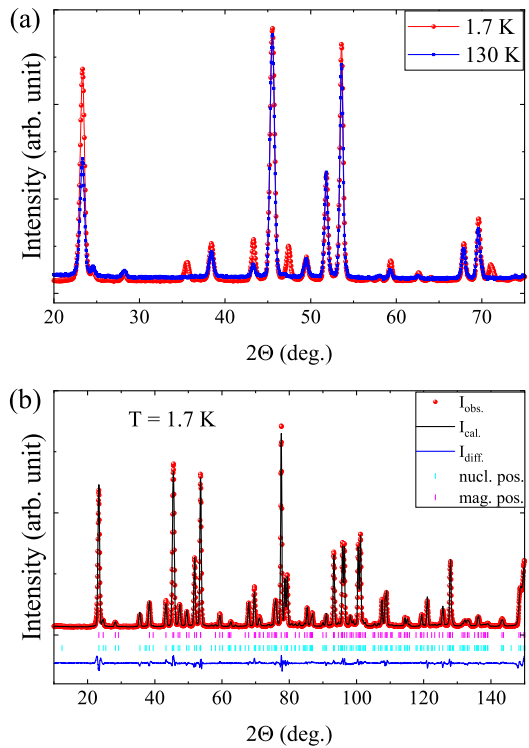}
\caption{(Color online) (a) Comparison of the neutron powder diffraction patterns measured above and
below the transition temperature. A neutron wavelength of 1.89 \AA ~was selected for the
measurements. (b) Rietveld magnetic structure refinement against the neutron pattern according to
irreducible representation $\Gamma_1$. The best fit yields $R_{mag}$ = 5.70\%.}
\label{neutron_pattern}
\end{figure}

\begin{figure}
\centering
\includegraphics[width=0.9\columnwidth]{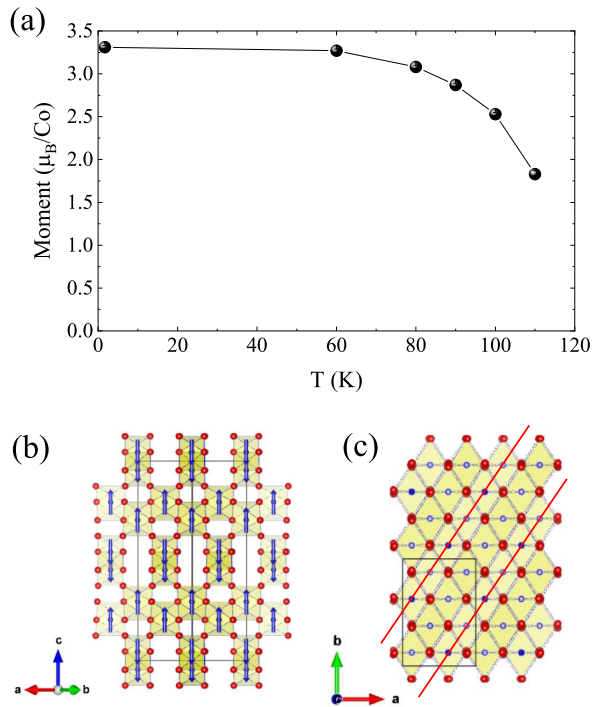}
\caption{(Color online) (a) Temperature dependence of the Co moments extracted from Rietveld
refinement. (b) and (c) The magnetic structure looked from different directions. In (c), the red
lines indicate possible missing layers due to the stacking faults. The blue and white circles
represent the spins pointing along the +\textbf{c} and -\textbf{c} directions, respectively.}
\label{magpattern}
\end{figure}

More insights to the magnetic ground state of this compound are obtained from neutron powder
diffraction measurements. Figure \ref{neutron_pattern}(a) compares the neutron pattern measured just
above the transition temperature and at the base temperature. Substantial magnetic intensities can
be observed on top of the nuclear reflections, indicating a \textbf{k} = 0 propagation vector. No
qualitative difference is observed in the intermediate temperature range (data not shown). The
magnetic structure was determined with the aid of irreducible representation analysis using the
\textit{BasIreps} program. For the Co ions at the \textit{16g} site with space group \textit{Fddd}
and propagation vector \textbf{k} = 0, the magnetic reducible representation is decomposed into
eight irreducible representations (IRs). The basis vectors for each IR can be found in SM Tab. S1.
The complexity due to stacking faults, as will be discussed later, and mixing valence state of Co
ions are neglected during the refinement. Therefore, the obtained magnetic structure should be
considered as an averaged one. Finally, it was found that only the IR $\Gamma_1$ can describe the
pattern satisfactorily.  The refined pattern is shown in Fig. \ref{neutron_pattern}, and the
temperature dependence of the moment size of Co ion is presented in Fig. \ref{magpattern}(a). The Co
moments are pointing either parallel or antiparallel to the \textit{c}-axis. The moments along the
zig-zag chains are coupled ferromagnetically, while the interchain moments are coupled
antiferromagnetically, following the Goodenough-Kanamori rules \cite{Kanamori1959}. Note that the
moment size at the base temperature amounts to 3.31 $\mu_B$/Co, which is larger than the spin only
value for the high spin state of Co$^{2+}$ (3 $\mu_B$), consistent with the expectation that
substantial orbital moments are not quenched from the high temperature Curie-Weiss fitting.

\section{Discussions}

The most striking observation of the orthorhombic Li$_3$Co$_2$SbO$_6$ is the multiple transitions
below $T_N$ from the dc and ac susceptibility measurements. These transitions have the following
characters: 1) they are ferromagnetic, although the overall ground state is antiferromagnetic; 2)
they are quite soft and can be easily suppressed by a modest magnetic field; 3) they are not
detected by other measurements such as heat capacity, $\mu$SR and neutron diffraction.
These results suggest that these transitions originate from a minor part of the sample and are
reminiscent of transitions seen in the layered manganites La$_{2-2x}$Sr$_{1+2x}$Mn$_2$O$_7$
(\textit{x} = 0.4), for which the ground state is ferromagnetic with $T_c \sim$ 125 K, but
additional ferromagnetic feature was observed near room temperature, that was also suppressed, or
masked, by an applied magnetic field \cite{Seshadri1997}, and was attributed to stacking faults of
the perovskite layers \cite{Moreno2000} and other members of the Ruddlesden-Popper series. These
peaks are also reminiscent of the multiple transitions observed in the $\alpha-$RuCl$_3$ due to
stacking faults of the honeycomb layers \cite{Yamauchi2018}.
Such a stacking faults scenario is also plausible for the Li$_3$Co$_2$SbO$_6$ case. From Fig.
\ref{magpattern}(c), one sees that the close-packed anion layered are stacked either along the [110]
or the [1-10] direction, and the Co ions are sandwiched between these layers. If some of the layers
as indicated by the red lines are missing, ferromagnetic regions could appear.

There could be other possibilities such as antisite disorder between the Li and Co ions as proposed
by Brown et al. \cite{Brown2019}, or the coexistence of Co$^{2+}$ and Co$^{3+}$ ions as found by our
refinement. In both cases, the appearance of Li or Co$^{3+}$ ions at the Co$^{2+}$ site could
disturb the long range magnetic ordering. For example, as shown in Fig. \ref{magpattern}(b), the
zig-zag chains running along the [110] and [1-10] directions are connected by corner-shared CoO6
octahedra, which favors antiferromagnetic interaction between the chains. If one of the Co$^{2+}$
ions mediating this inerchain interaction is missing, the local structure around this defect may
become ferromagnetic. Finally, the real situation could be a complicate combination of these
individual effects.

The stacking faults or local site disorders may be inferred from the $\mu$SR results. In Fig.
\ref{muSR_parameter}(a), the relaxation rate, which is proportional to the internal field
distribution width, corresponding to the internal field $B_2$ is much larger than that of $B_1$. It
can be surmised that muons that stopped around the stacking faults or local disorders contribute to
the $B_2$ component, while those far from the defects contribute to the $B_1$ component, as was
observed in Na$_2$IrO$_3$ \cite{Choi2012}. Note that the amplitude of $A_1$ and $A_2$ are
comparable, which seems to contradict with the conjecture that the ferromagnetic component only have
a minor volume fraction. The argument is that the muons feel the field distribution via long range
dipolar field, which extends over the ferromagnetic boundaries. On the other hand, the possibility
that there exists two inequivalent muon sites with similar population cannot be ruled out, future
DFT calculations will be helpful to clarify this point.

Another enigmatic phenomenon is the distinct transition temperature observed by ac and dc magnetic
susceptibility measurements. One possibility is the different amplitude of the magnetic fields
applied during the measurement, and the pinning barriers for the domains are too high for the ac
measurement, as observed in the ferrimagnetic system FeCr$_2$S$_4$ \cite{Tsurkan-2001}.
Nevertheless, unexplained differences between static and dynamic susceptibility are rare, and the
origins are still open.

\section{Conclusions}

The magnetic properties of the orthorhombic Li$_3$Co$_2$SbO$_6$ have been studied in detail by means
of dc and ac susceptibility, muon spin relaxation and neutron diffraction measurements. The sample
forms an antiferromagetic ground state, with ferromagnetic chains stacked antiferromagnetically
along the \textit{c} axis. Additional transitions below $T_N$ are observed by dc and ac magnetic
susceptibility measurements, which are ascribed to possible stacking faults and/or local disorders,
resulting in the ferromagnetic component within the overall antiferromagnetic matrix. However,
single crystal studies, probing local magnetic and crystallographic structure are required to fully
understand the magnetic response in orthorhombic Li$_3$Co$_2$SbO$_6$. Those findings are also
suggestive for the honeycomb phase in which the honeycomb layers are prone to stacking faults,
either nuclear structurally or magnetically, which may not be readily detected by bulk structural
measurements.

H. G. acknowledges the support from the NSF of China with Grant No. 12004270, and Guangdong Basic
and Applied Basic Research Foundation (2019A1515110517). This research is also partly supported by
the NSF of China with Grant No. U1932155. A portion of this work was supported by the Laboratory
Directed Research and Development (LDRD) program of Oak Ridge National Laboratory, managed by
UT-Battelle, LLC for the U.S. Department of Energy.


\bibliography{LCSO}

\section{Supporting Information}
\setcounter{figure}{0}
\setcounter{table}{0}
\renewcommand{\thefigure}{S\arabic{figure}}
\begin{figure}[b]
\centering
\includegraphics[width=0.8\columnwidth]{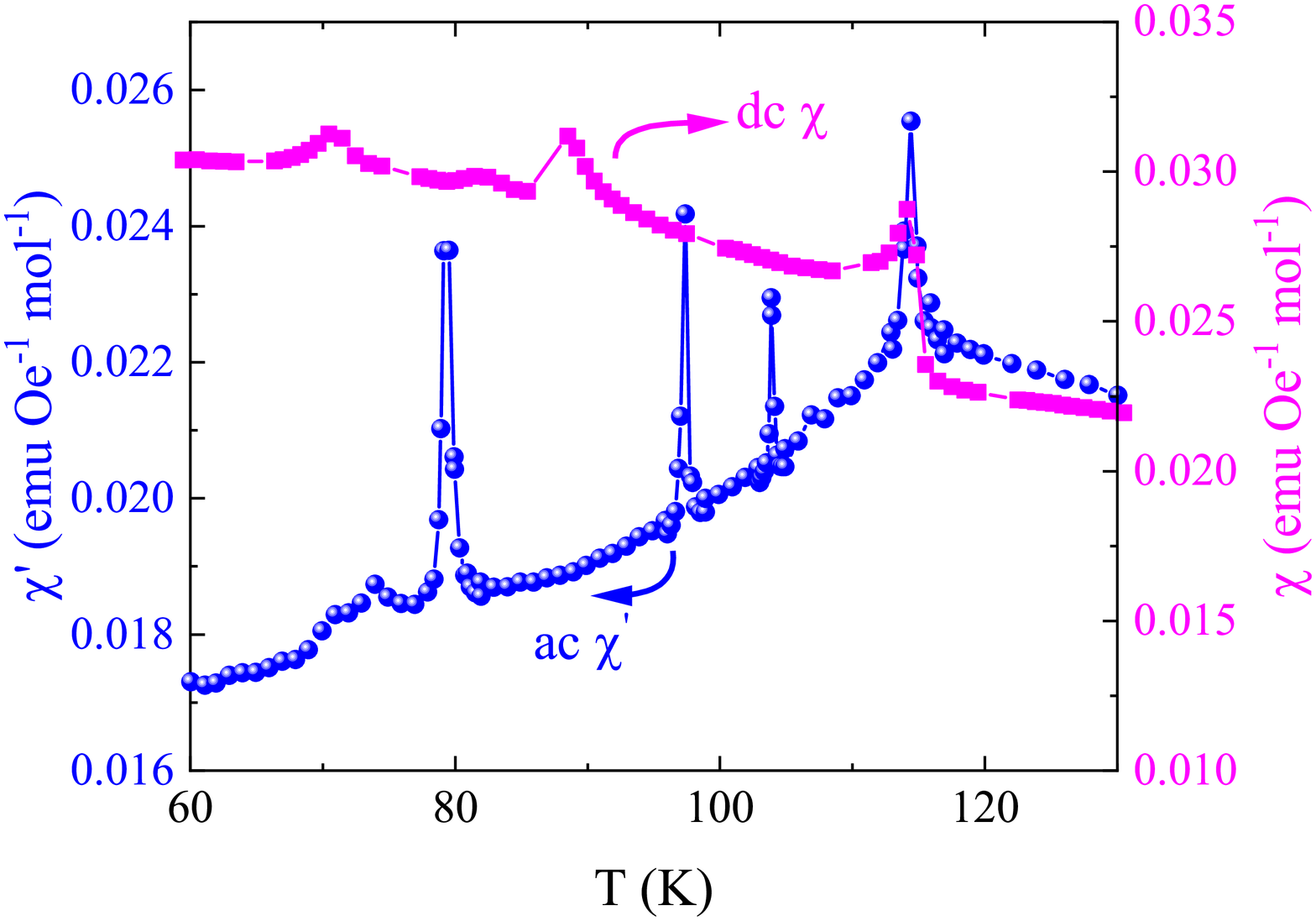}
\caption{(Color online) Comparison of the temperature dependence of the dc and ac magnetic
susceptibility measured exactly for the same sample.}
\label{mt}
\end{figure}
Figure \ref{mt} shows additional magnetic susceptibility measurements exactly on the same sample
using the AC and DC mode of the ACMS option in the PPMS. The dc susceptibility was measured in the
zero-field-cooled mode, and a magnetic field of 0.1 T was applied during the measurement. The ac
susceptibility was measured at 3782 Hz with the driving field of 5 Oe. These results are similar to
the ones shown in the main text, with sharp peaks, e.g. at 79 and 97 and 104 K, only appearing in
the ac susceptibility. Some missing points in the dc susceptibility is due to touchdown operation
during the measurements.
The magnetic symmetry has been analyzed by the irreducible representation (IR) theory using the
BasIrep program within the FullProf program suite. The basis vectors for each IR are listed in Tab.
\ref{BV}, together with the $R_{mag}$ factor obtained by the Rietveld refinement.

\renewcommand{\thetable}{S\arabic{table}}
\begin{table}
\caption{Irreducible representations (IR) and the basis vectors $\varphi$ for the Co ions at the
16\textit{g} site with the  space group \textit{Fddd} (No. 70) and propagation vector
\textbf{k}~=~(0 0 0). Site1 - (x, y, z), site2 - (-x+3/4, y, -z+3/4), site3 - (-x+1/2, -y+1/2, -z+1)
and site4 - (x+1/4, -y+1, z+1/4 ).\label{BV}}
\begin{ruledtabular}
\begin{tabular}{lcccccc}
IRs        &   $R_{mag}$ (\%)  & $\varphi$  &  site1 & site2 & site3  & site4\\
\hline
$\Gamma_1$ &  5.70  &  $\varphi_1$  & (0~0~1) & (0~0~-1) & (0~0~1) & (0~0~-1) \\
$\Gamma_2$ &  60.9  &  $\varphi_1$  & (0~0~1) & (0~0~-1) & (0~0~-1)& (0~0~1) \\
$\Gamma_3$ &  69.3  &  $\varphi_1$  & (0~0~1) & (0~0~1) & (0~0~1) & (0~0~1) \\
$\Gamma_4$ &  88.3  &  $\varphi_1$  & (0~0~1) &(0~0~1) & (0~0~-1) & (0~0~-1) \\
$\Gamma_5$ &  72.1  &  $\varphi_1$  & (1~0~0) & (-1~0~0) & (1~0~0) & (-1~0~0) \\
           &        &  $\varphi_2$  & (0~1~0) & (0~1~0) & (0~1~0) & (0~1~0)  \\
$\Gamma_6$ &  63.2  &  $\varphi_1$  & (1~0~0) & (-1~0~0) & (-1~0~0) & (1~0~0)  \\
           &        &  $\varphi_2$  & (0~1~0) & (0~1~0) & (0~-1~0) & (0~-1~0)  \\
$\Gamma_7$ &  69.8  &  $\varphi_1$  & (1~0~0) & (1~0~0) & (1~0~0) & (1~0~0)  \\
           &        &  $\varphi_2$  & (0~1~0) & (0~-1~0) & (0~1~0) & (0~-1~0) \\
$\Gamma_8$ &  56.5  &  $\varphi_1$  & (1~0~0) & (1~0~0) & (-1~0~0) & (-1~0~0) \\
           &        &  $\varphi_2$  & (0~1~0) & (0~-1~0) & (0~-1~0) & (0~1~0) \\
\end{tabular}
\end{ruledtabular}
\end{table}

\end{document}